\documentstyle[12pt]{article}
\newcommand{\cl}{\centerline}

\newcommand\beq{\begin{equation}}
\newcommand\eeq{\end{equation}}
\newcommand\bea{\begin{eqnarray}}
\newcommand\eea{\end{eqnarray}}

\begin{document}

\begin{titlepage}
\setlength{\textwidth}{5.0in} \setlength{\textheight}{7.5in}
\setlength{\parskip}{0.0in} \setlength{\baselineskip}{18.2pt}
\hfill {\tt SOGANG-HEP 262/99}
\begin{center}
\cl{\Large{{\bf BFT Hamiltonian embedding for SU(3) Skyrmion}}}\par
\vskip 0.5cm
\begin{center}
{Soon-Tae Hong$^a$ and Young-Jai Park$^{b}$}\par
\end{center}
\vskip 0.4cm
\begin{center}
{Department of Physics and Basic Science Research Institute,}\par
{Sogang University, C.P.O. Box 1142, Seoul 100-611, Korea}\par
\end{center}
\vskip 0.5cm
\cl{\today}
\vfill
\begin{center}
{\bf ABSTRACT}
\end{center}
\begin{quotation}

We newly apply the Batalin, Fradkin and Tyutin (BFT) formalism to the
SU(3) flavor Skyrmion model to investigate the Weyl ordering
correction to the structure of the hyperfine splittings of strange
baryons.  On the other hand, the Berry phases and Casimir effects are 
also discussed.   

\vskip 0.5cm \noindent PACS: 12.39.Dc, 11.10.Ef, 14.20.-c\\
\noindent Keywords: Skyrmion, Collective Coordinates, BFT
formalism
\par
---------------------------------------------------------------------\\
\noindent $^a$sthong@ccs.sogang.ac.kr\\ \noindent
$^b$yjpark@ccs.sogang.ac.kr \vskip 0.5cm \noindent
\end{quotation}
\end{center}
\end{titlepage}

Since the SU(2) Skyrme model \cite{sky} had been proposed, it is 
well known that baryons can be obtained from topological
solutions, known as SU(2) Skyrmions, because the homotopy group
$\Pi_{3}(SU(2))=Z$ admits fermions \cite{ad,sk,hsk}. Using the
collective coordinates of the isospin rotation of the Skyrmion,
Adkins et al. \cite{ad} have performed semiclassical quantization 
having the static properties of baryons within 30\% of the 
corresponding experimental data.  After their work, several authors 
have suggested interesting schemes \cite{sk,hsk,yabu,callan,sco,
kleb94,kleb96,ike99,multi} to improve the predictions 
of physical observables for baryons.  Furthermore, the hyperfine splittings 
for the SU(3) phenomenological Skyrmion have been studied in two main 
schemes.  Firstly, the SU(3) 
cranking method exploits rigid rotation of the Skyrmion in the 
collective space of SU(3) Euler angles with full diagonalization of 
the flavor-symmetry breaking terms.  Especially, Yabu and Ando \cite{yabu} 
proposed the exactly diagonalization of the symmetry breaking terms by 
introducing the higher irreducible representation mixing in the baryon 
wave function, which was later interpreted in terms of the multiquark 
structure \cite{multi} in the baryon wave function.  Secondly, Callan 
and Klebanov \cite{callan} suggested an interpretation of baryons 
containing a heavy quark as bound states of solitons of the pion chiral 
lagrangian with mesons.  In their formalism, the fluctuations in the 
strangeness direction are treated differently from those in the isospin 
directions \cite{callan,sco}.  Recently, Klebanov and Westerberg 
\cite{kleb94,kleb96} 
proposed rigid rotator approach to the SU(3) Skyrmion, where the rigid 
motions of the SU(3) Skyrmion are separated into the SU(2) rotations 
and the deviations into strange directions to yield the structure of 
the hyperfine splittings of strange baryons.

On the other hand, quite recently it has been shown that a Hamiltonian 
embedding method \cite{BFT} developed by Batalin, Fradkin and Tyutin (BFT), 
which converts the second class constraints into first class ones by 
introducing auxiliary fields, gives an additional energy term in 
SU(2) Skyrmion model\cite{hkp98}.  We have also analyzed the symmetric 
structure of the corresponding Lagrangian \cite{hkp99}.

The motivation of this paper is to extend the BFT scheme for the
SU(2) Skyrmion to the SU(3) flavor case so that one can
investigate the Weyl ordering correction to $c$ the ratio of the
strange-light to light-light interaction strengths and $\bar{c}$
that of the strange-strange to light-light.  Differently from the
Klebanov and Westerberg's standard rigid rotator
approach\cite{kleb94} to the SU(3) Skyrmion where the angular
velocity of the SU(2) rotation was used, we exploit the SU(2)
collective coordinates which are naturally embedded in the SU(3)
group manifold so that, as in the SU(2) flavor case, one can
introduce the BFT scheme in the SU(3) Skyrmion to yield the
modified baryon energy spectrum and the structure of the hyperfine
splittings of strange baryons.

Now we start with the SU(3) Skyrmion Lagrangian of the form
\begin{eqnarray}
L&=&\int{\rm d}^{3}r[\frac{f_{\pi}^{2}}{4}{\rm tr}(\partial_{\mu}U^{\dag}%
\partial^{\mu}U)+\frac{1} {32e^{2}}{\rm tr}[U^{\dag}\partial_{\mu}U,U^{\dag}%
\partial_{\nu}U]^{2}  \nonumber \\
& &+\frac{f_{\pi}^{2}}{4}{\rm tr}M(U+U^{\dag}-2)]+L_{WZW},
\end{eqnarray}
where $f_{\pi}$ is the pion decay constant and $e$ is a
dimensionless parameter, and $U$ is an SU(3) matrix. $M$ is
proportional to the quark mass matrix given by
\[
M={\rm diag}~ (m_{\pi}^{2},~ m_{\pi}^{2},~ 2m_{K}^{2}-m_{\pi}^{2})
\label{m}
\]
where $m_{\pi}=138$ MeV, which is often neglected, and $m_{K}=495$
MeV.  The Wess-Zumino-Witten (WZW) term\cite{wzw} is described by
the action $$
\Gamma_{WZW}=-\frac{iN}{240\pi^{2}}\int_{{\sf M}}{\rm d}^{5}r\epsilon^{\mu%
\nu \alpha\beta\gamma}{\rm
tr}(l_{\mu}l_{\nu}l_{\alpha}l_{\beta}l_{\gamma}) $$
where $N$ is the number of colors,
$l_{\mu}=U^{\dag}\partial_{\mu}U$ and the integral is done on the
five-dimensional manifold ${\sf M}=V\times S^{1}\times I$ with the
three-space volume $V$, the compactified time $S^{1}$ and the unit
interval $I$ needed for a local form of WZW term.

Assuming maximal symmetry in the Skyrmion, we describe the
hedgehog solution $U_{0}$ embedded in the SU(2) isospin subgroup
of SU(3)
\[
U_{0}(\vec{x})=\left(
\begin{array}{cc}
e^{i\vec{\tau}\cdot\hat{x}f(r)} & 0 \\ 0 & 1
\end{array}
\right) \label{u}
\]
where the $\tau_{i}$ ($i$=1,2,3) are Pauli matrices,
$\hat{x}=\vec{x}/r$ and $f(r)$ is the chiral angle determined by
minimizing the static mass $E$ given below and for unit winding
number $\lim_{r \rightarrow \infty} f(r)=0$ and $f(0)=\pi$. On the
other hand, since the hedgehog ansatz has maximal or spherical
symmetry, it is easily seen that spin plus isospin equals zero, so
that isospin transformations and spatial rotations are related to
each other.

Now we consider only the rigid motions of the SU(3) Skyrmion $$
U(\vec{x},t)={\cal A}(t)U_{0}(\vec{x}){\cal A}(t)^{\dag}
\label{uxt} $$
where, to separate the SU(2) rotations from the deviations into
strange directions, the time-dependent rotations can be written
as\cite{kleb90}
\[
{\cal A}(t)=\left(
\begin{array}{cc}
A(t) & 0 \\ 0 & 1
\end{array}
\right)S(t) 
\]
with $A(t) \in$ SU(2) and the small rigid oscillations $S(t)$
around the SU(2) rotations\cite{kleb90}.  Furthermore, in the
SU(2) subgroup of SU(3), the spin and isospin states can be
treated by the time-dependent collective coordinates
$a^{\mu}=(a^{0},\vec{a})$ $(\mu=0,1,2,3)$ corresponding to the
spin and isospin rotations as in the standard SU(2) Skyrmion
\[
A(t) = a^{0}+i\vec{a}\cdot\vec{\tau}.
\]
With the hedgehog ansatz and the collective rotation $A(t)\in$
SU(2) in the
SU(2) embedding in the SU(3) manifold, the chiral field can be given by $U(%
\vec{x},t)=A(t)U_{0}(\vec{x})
A^{\dagger}(t)=e^{i\tau_{a}R_{ab}\hat{x}_{b} f(r)}$ where
$R_{ab}=\frac{1}{2} {\rm tr} (\tau_{a}A\tau_{b}A^{\dagger})$.

On the other hand the small rigid oscillations $S$, which were
also used in Ref. \cite{kleb94}, can be described as
\[
S(t)={\rm exp}(i\sum_{a=4}^{7}d^{a}\lambda_{a})={\rm exp}(i{\cal
D}),
\]
where
\[
{\cal D}=\left(
\begin{array}{cc}
0 & \sqrt{2}D \\ \sqrt{2}D^{\dag} & 0
\end{array}
\right),~~ D=\frac{1}{\sqrt{2}}\left(
\begin{array}{c}
d^{4}-id^{5} \\ d^{6}-id^{7}
\end{array}
\right). 
\]

After some algebra, the Skyrmion Lagrangian to order $1/N$ is then
given in terms of the SU(2) collective coordinates $a^{\mu}$ and
the strange deviations $D$
\begin{eqnarray}
L&=&-E-\frac{1}{2}\chi m_{\pi}^{2}+2{\cal
I}_{1}\dot{a}^{\mu}\dot{a}^{\mu} +4{\cal
I}_{2}\dot{D}^{\dag}\dot{D}+\frac{i}{2}N(D^{\dag}\dot{D}
-\dot{D}^{\dag}D)\nonumber\\ & &-\chi
(m_{K}^{2}-m_{\pi}^{2})D^{\dag}D
+2i({\cal I}_{1}-2{\cal I}_{2})\{D^{\dag} (a^{0}\vec{\dot{a}}-\dot{a}^{0}%
\vec{a}+\vec{a}\times\vec{\dot{a}}) \cdot\vec{\tau}\dot{D}
\nonumber \\
& &-\dot{D}^{\dag}(a^{0}\vec{\dot{a}}-\dot{a}^{0}\vec{a} +\vec{a}\times\vec{%
\dot{a}})\cdot\vec{\tau}D\} -ND^{\dag}(a^{0}\vec{\dot{a}}-\dot{a}^{0}\vec{a}+%
\vec{a}\times\vec{\dot{a}}) \cdot\vec{\tau}D  \nonumber \\
& &+2({\cal I}_{1}-\frac{4}{3}{\cal I}_{2})(D^{\dag}D)(\dot{D}^{\dag}\dot{D}%
) -\frac{1}{2}({\cal I}_{1}-\frac{4}{3}{\cal
I}_{2})(D^{\dag}\dot{D}+\dot{D} ^{\dag}D)^{2}  \nonumber \\
& &+2{\cal I}_{2}(D^{\dag}\dot{D}-\dot{D}^{\dag}D)^{2} -\frac{i}{3}N(D^{\dag}%
\dot{D}-\dot{D}^{\dag}D)D^{\dag}D\nonumber\\ & &+\frac{2}{3}\chi
(m_{K}^{2}-m_{\pi}^{2})(D^{\dag}D)^{2} \label{lag}
\end{eqnarray}
where the soliton energy $E$, the moments of inertia ${\cal I}_{1}$ and $%
{\cal I}_{2}$, and the strength $\chi$ of the chiral symmetry
breaking are respectively given by
\begin{eqnarray}
E&=&4\pi\int_{0}^{\infty}{\rm d}r r^{2}[\frac{f_{\pi}^{2}}{2}((\frac{{\rm d}f%
} {{\rm d}r})^ {2}+\frac{2\sin^{2}f}{r^{2}}) +\frac{1}{2e^{2}}\frac{\sin^{2}f%
}{r^{2}} (2(\frac{{\rm d}f}{{\rm
d}r})^{2}+\frac{\sin^{2}f}{r^{2}})], \nonumber \\ {\cal
I}_{1}&=&\frac{8\pi}{3}\int_{0}^{\infty}{\rm d}r
r^{2}\sin^{2}f[f_{\pi}^{2} +\frac{1}{e^{2}}((\frac{{\rm d}f}{{\rm d}r})^{2}+%
\frac{\sin^{2}f}{r^{2}})]  \nonumber \\
{\cal I}_{2}&=&2\pi\int_{0}^{\infty}{\rm d}r r^{2}(1-\cos f)[f_{\pi}^{2} +%
\frac{1}{4e^{2}}((\frac{{\rm d}f}{{\rm
d}r})^{2}+\frac{2\sin^{2}f}{r^{2}})] \nonumber \\ \chi&=&8\pi
f_{\pi}^{2}\int_{0}^{\infty}{\rm d}r r^{2}(1-\cos f). \label{eni}
\end{eqnarray}
The momenta $\pi^{\mu}$ and $\pi_{s}^{\alpha}$, conjugate to the
collective coordinates $a^{\mu}$ and the strange deviation
$D_{\alpha}^{\dag}$ are given by
\begin{eqnarray}
\pi^{0}&=&4{\cal I}_{1}\dot{a}^{0}-2i({\cal I}_{1}-2{\cal I}_{2})
(D^{\dag}\vec{a}\cdot\vec{\tau}\dot{D}-\dot{D}^{\dag}\vec{a}\cdot\vec{\tau}%
D) +ND^{\dag}\vec{a}\cdot\vec{\tau}D  \nonumber \\
\vec{\pi}&=&4{\cal I}_{1}\vec{\dot{a}}+2i({\cal I}_{1}-2{\cal
I}_{2})
\{D^{\dag}(a^{0}\vec{\tau}-\vec{a}\times\vec{\tau})\dot{D} -\dot{D}%
^{\dag}(a^{0}\vec{\tau}-\vec{a}\times\vec{\tau})D\}  \nonumber \\
& &-ND^{\dag}(a^{0}\vec{\tau}-\vec{a}\times\vec{\tau})D  \nonumber
\\ \pi_{s}&=&4{\cal I}_{2}\dot{D}-\frac{i}{2}ND-2i({\cal
I}_{1}-2{\cal I}_{2})
(a^{0}\vec{\dot{a}}-\dot{a}^{0}\vec{a}+\vec{a}\times\vec{\dot{a}}) \cdot\vec{%
\tau}D  \nonumber \\
& &+2({\cal I}_{1}-\frac{4}{3}{\cal I}_{2})(D^{\dag}D)\dot{D} -({\cal I}_{1}-%
\frac{4}{3}{\cal I}_{2})(D^{\dag}\dot{D}+\dot{D}^{\dag}D)D
\nonumber \\ & &-4{\cal
I}_{2}(D^{\dag}\dot{D}-\dot{D}^{\dag}D)D+\frac{i}{3}N(D^{\dag}D)D
\nonumber
\end{eqnarray}
which satisfy the Poisson brackets $$
\{a^{\mu},\pi^{\nu}\}=\delta^{\mu\nu},~~~\{D_{\alpha}^{\dag},
\pi_{s}^{\beta}\}=\{D^{\beta},\pi_{s,\alpha}^{\dag}\}=\delta_{\alpha}^{\beta}.
$$

Performing Legendre transformation, we obtain the Hamiltonian to
order $1/N$ as follows
\begin{eqnarray}
H&=&E+\frac{1}{2}\chi m_{\pi}^{2}+\frac{1}{8{\cal
I}_{1}}\pi^{\mu}\pi^{\mu} +\frac{1}{4{\cal
I}_{2}}\pi_{s}^{\dag}\pi_{s}-i\frac{N}{8{\cal I}_{2}}
(D^{\dag}\pi_{s}-\pi_{s}^{\dag}D) +(\frac{N^{2}}{16{\cal
I}_{2}}\nonumber\\ & &+\chi
(m_{K}^{2}-m_{\pi}^{2}))D^{\dag}D+i(\frac{1}{4{\cal I}_{1}}
-\frac{1}{8{\cal I}_{2}}) \{D^{\dag}
(a^{0}\vec{\pi}-\vec{a}\pi^{0}+\vec{a}\times\vec{\pi})
\cdot\vec{\tau}\pi_{s}  \nonumber
\\
& &-\pi_{s}^{\dag}(a^{0}\vec{\pi}-\vec{a}\pi^{0} +\vec{a}\times\vec{\pi}%
)\cdot\vec{\tau}D\} +\frac{N}{8{\cal I}_{2}}D^{\dag}(a^{0}\vec{\pi}-\vec{a}%
\pi^{0} +\vec{a}\times\vec{\pi})\cdot\vec{\tau}D  \nonumber \\ &
&+(\frac{1}{2{\cal I}_{1}}-\frac{1}{3{\cal I}_{2}})(D^{\dag}D)
(\pi_{s}^{\dag}\pi_{s})+(\frac{1}{12{\cal I}_{2}}-\frac{1}{8{\cal
I}_{1}}) (D^{\dag}\pi_{s}+\pi_{s}^{\dag}D)^{2}  \nonumber \\
& &-\frac{1}{8{\cal I}_{2}}(D^{\dag}\pi_{s}-\pi_{s}^{\dag}D)^{2} -i\frac{N}{8%
{\cal I}_{2}}(D^{\dag}\pi_{s}-\pi_{s}^{\dag}D)(D^{\dag}D)
\nonumber \\ & &+(\frac{N^{2}}{12{\cal I}_{2}}-\frac{2}{3}\chi
(m_{K}^{2}-m_{\pi}^{2})) (D^{\dag}D)^{2}. \label{hamil}
\end{eqnarray}

On the other hand, since the physical fields $a^{\mu}$ are the
collective
coordinates of the SU(2) group manifold isomorphic to the hypersphere $S^{3}$%
, we have the second class constraints
\begin{eqnarray}
\Omega_{1} &=& a^{\mu}a^{\mu}-1\approx 0,  \nonumber \\ \Omega_{2}
&=& a^{\mu}\pi^{\mu}\approx 0.  \label{omega2}
\end{eqnarray}
Here one notes that the derivation of the second constraint is not
trivial differently from that in the SU(2) Skyrmion
model\cite{hkp98} where the constraints (\ref{omega2}) also hold.
In other words, through the following complicated algebraic
relations
\begin{eqnarray}
\{a^{0}, H\}&=&\frac{1}{4{\cal I}_{1}}\pi^{0}-i(\frac{1}{4{\cal I}_{1}} -%
\frac{1}{8{\cal I}_{2}})(D^{\dag}\vec{a}\cdot\vec{\tau}\pi_{s}
-\pi_{s}^{\dag}\vec{a}\cdot\vec{\tau}D)-\frac{N}{8{\cal I}_{2}}D^{\dag} \vec{%
a}\cdot\vec{\tau}D  \nonumber \\
\{\vec{a}, H\}&=&\frac{1}{4{\cal I}_{1}}\vec{\pi}+i(\frac{1}{4{\cal I}%
_{1}} -\frac{1}{8{\cal I}_{2}})\{D^{\dag}(a^{0}\vec{\tau}-\vec{a}\times\vec{%
\tau}) \pi_{s}  \nonumber \\
& &-\pi_{s}^{\dag}(a^{0}\vec{\tau}-\vec{a}\times\vec{\tau})D\} +\frac{N}{8%
{\cal I}_{2}}D^{\dag}(a^{0}\vec{\tau}-\vec{a}\times\vec{\tau})D,
\nonumber
\end{eqnarray}
we can obtain the Poisson commutator
\[
\{\Omega_1,H\}={\frac{1}{2{\cal {I}}}}\Omega_{2}
\]
which yields the second constraint of (\ref{omega2}) since the
secondary constraint comes from the time evolution of
$\Omega_{1}$. The above two constraints then yield the Poisson
algebra $$
\Delta_{k k^{\prime}}=\{\Omega_{k},\Omega_{k^{\prime}}\} =
2\epsilon^{k k^{\prime}}a^{\mu}a^{\mu}  \label{delta} $$
with $\epsilon^{12}=-\epsilon^{21}=1$. Using the Dirac bracket
\cite{di} defined by
\[
\{A,B\}_{D}=\{A,B\}-\{A,\Omega_{k}\}\Delta^{k k^{\prime}}
\{\Omega_{k^{\prime}},B\}
\]
with $\Delta^{k k^{\prime}}$ being the inverse of $\Delta_{k
k^{\prime}}$ one can obtain the commutator relations
\begin{eqnarray}
\{a^{\mu},a^{\nu}\}_{D}&=&0,  \nonumber \\
\{a^{\mu},\pi^{\nu}\}_{D}&=&\delta^{\mu \nu}-\frac{a^{\mu}a^{\nu}}{%
a^{\sigma} a^{\sigma}},  \nonumber \\
\{\pi^{\mu},\pi^{\nu}\}_{D}&=&\frac{1}{a^{\sigma}a^{\sigma}}
(a^{\nu}\pi^{\mu}-a^{\mu}\pi^{\nu}). \nonumber
\end{eqnarray}

Now, maintaining the SU(2) symmetry originated from the massless
$u$- and $d$-quarks and following the abelian BFT formalism
\cite{BFT,hkp98} which systematically converts the second class
constraints into first class ones, we introduce two auxiliary
fields $\Phi^{i}$ corresponding to $\Omega_{i}$ with the Poisson
brackets $$
\{\Phi^{i}, \Phi^{j}\}=\epsilon^{ij}. $$
The first class constraints $\tilde{\Omega}_{i}$ are then
constructed as a power series of the auxiliary fields:
\begin{equation}
\tilde{\Omega}_{i}=\sum_{n=0}^{\infty}\Omega_{i}^{(n)},~~~~
\Omega_{i}^{(0)}=\Omega_{i}  \label{tilin}
\end{equation}
where $\Omega_{i}^{(n)}$ are polynomials in the auxiliary fields
$\Phi^{j}$ of degree $n$, to be determined by the requirement that
the first class constraints $\tilde{\Omega}_{i}$ satisfy an
abelian algebra as follows
\begin{equation}
\{\tilde{\Omega}_{i},\tilde{\Omega}_{j}\}=0.  \label{cijk}
\end{equation}
After some algebraic manipulations, one can obtain the desired
first class constraints
\begin{eqnarray}
\tilde{\Omega}_{1}&=&\Omega_{1}+2\Phi^{1},  \nonumber \\
\tilde{\Omega}_{2}&=&\Omega_{2}-a^{\mu}a^{\mu}\Phi^{2} \nonumber
\end{eqnarray}
which yield the strongly involutive first class constraint algebra
(\ref {cijk}). On the other hand, the corresponding first class
Hamiltonian is given by
\begin{eqnarray}
\tilde{H}&=&E+\frac{1}{2}\chi m_{\pi}^{2}+\frac{1}{8{\cal I}_{1}}
(\pi^{\mu}-a^{\mu}\Phi^{2})
(\pi^{\mu}-a^{\mu}\Phi^{2})\frac{a^{\nu}a^{\nu}}{a^{\nu}a^{\nu}+2\Phi^{1}%
}  \nonumber \\
& &+\frac{1}{4{\cal I}_{2}} \pi_{s}^{\dag}\pi_{s}-i\frac{N}{8{\cal I}_{2}}%
(D^{\dag}\pi_{s}-\pi_{s}^{\dag}D) +(\frac{N^{2}}{16{\cal
I}_{2}}+\chi (m_{K}^{2}-m_{\pi}^{2}))D^{\dag}D  \nonumber \\
& &+i(\frac{1}{4{\cal I}_{1}}-\frac{1}{8{\cal I}_{2}})\{D^{\dag} (a^{0}\vec{%
\pi}-\vec{a}\pi^{0}+\vec{a}\times\vec{\pi}) \cdot\vec{\tau}%
\pi_{s}  \nonumber \\
& &-\pi_{s}^{\dag}(a^{0}\vec{\pi}-\vec{a}\pi^{0} +\vec{a}\times\vec{%
\pi})\cdot\vec{\tau}D\} +\frac{N}{8{\cal I}_{2}}D^{\dag}(a^{0}\vec{\pi}%
-\vec{a}\pi^{0} +\vec{a}\times\vec{\pi})\cdot\vec{\tau}D \nonumber
\\ & &+\cdots
\label{hct}
\end{eqnarray}
where the ellipsis stands for the strange-strange interaction
terms of order $1/N$ which can be readily read off from Eq.
(\ref{hamil}).  Here one notes that the BFT corrections for the
second class constraints do not affect even the isospin-strange
coupling terms with the Pauli matrices $\tau_{i}$.  The above
first class Hamiltonian is also strongly involutive with the first
class constraints
\[
\{\tilde{\Omega}_{i},\tilde{H}\}=0.
\]

Now the substitution of the first class constraints into the
Hamiltonian (\ref{hct}) yields the Hamiltonian of the form \bea
\tilde{H}&=&E+\frac{1}{2}\chi m_{\pi}^{2}+\frac{1}{8{\cal I}_{1}}
(a^{\mu}a^{\mu}\pi^{\nu}\pi^{\nu}-a^{\mu}\pi^{\mu}a^{\nu}\pi^{\nu})\nonumber\\
& &+\frac{1}{4{\cal I}_{2}} \pi_{s}^{\dag}\pi_{s}-i\frac{N}{8{\cal I}_{2}}%
(D^{\dag}\pi_{s}-\pi_{s}^{\dag}D) +(\frac{N^{2}}{16{\cal
I}_{2}}+\chi (m_{K}^{2}-m_{\pi}^{2}))D^{\dag}D  \nonumber \\
& &+i(\frac{1}{4{\cal I}_{1}}-\frac{1}{8{\cal I}_{2}})\{D^{\dag} (a^{0}\vec{%
\pi}-\vec{a}\pi^{0}+\vec{a}\times\vec{\pi}) \cdot\vec{\tau}%
\pi_{s}  \nonumber \\
& &-\pi_{s}^{\dag}(a^{0}\vec{\pi}-\vec{a}\pi^{0} +\vec{a}\times\vec{%
\pi})\cdot\vec{\tau}D\} +\frac{N}{8{\cal I}_{2}}D^{\dag}(a^{0}\vec{\pi}%
-\vec{a}\pi^{0} +\vec{a}\times\vec{\pi})\cdot\vec{\tau}D \nonumber
\\ & &+\cdots. \label{h1st} \eea Following the symmetrization
procedure\cite{hkp98}, we obtain the Weyl ordering correction to
the first class Hamiltonian (\ref{h1st}) as follows \bea
\tilde{H}&=&E+\frac{1}{2}\chi m_{\pi}^{2}+\frac{1}{2{\cal I}_{1}}
(\vec{I}^{2}+\frac{1}{4})+\frac{1}{4{\cal I}_{2}}
\pi_{s}^{\dag}\pi_{s} -i\frac{N}{8{\cal
I}_{2}}(D^{\dag}\pi_{s}-\pi_{s}^{\dag}D)\nonumber\\ &
&+(\frac{N^{2}}{16{\cal I}_{2}}+\chi
(m_{K}^{2}-m_{\pi}^{2}))D^{\dag}D +i(\frac{1}{2{\cal
I}_{1}}-\frac{1}{4{\cal I}_{2}})(D^{\dag}\vec{I}\cdot
\vec{\tau}\pi_{s}-\pi_{s}^{\dag}\vec{I}\cdot\vec{\tau}D)\nonumber\\
& &+\frac{N}{4{\cal I}_{2}}D^{\dag}\vec{I}\cdot\vec{\tau}D+\cdots.
\label{nht} \eea where, as in the SU(2) standard Skyrmion, the
isospin operator $\vec{I}$ is given by\cite{ad} $$
\vec{I}=\frac{1}{2}(a^{0}\vec{\pi}-\vec{a}\pi^{0}
+\vec{a}\times\vec{\pi}) $$
which itself is invariant under the Weyl ordering procedure.
Here, by using the SU(2) collective coordinates $a^{\mu}$ instead
of the angular velocity of the SU(2) rotation
$A^{\dag}\dot{A}=\frac{i}{2}\dot{\vec{\alpha}}\cdot \vec{\tau}$
used in Ref.\cite{kleb94}, we have obtained the same result
(\ref{nht}) as that of Klebanov and Westerberg (KW) \cite{kleb94},
apart from the overall energy shift $\frac{1}{8{\cal I}_{1}}$
originated from the BFT correction.

Following the quantization scheme of KW for the strangeness flavor
direction, one can obtain the Hamiltonian of the form \bea
\tilde{H}&=&E+\frac{1}{2}\chi m_{\pi}^{2}+\frac{1}{2{\cal I}_{1}}
(\vec{I}^{2}+\frac{1}{4})+\frac{N}{8{\cal I}_{2}}(\mu -1)a^{\dag}a
\nonumber\\ & &+(\frac{1}{2{\cal I}_{1}} -\frac{1}{4{\cal
I}_{2}\mu}(\mu -1))a^{\dag}\vec{I} \cdot\vec{\tau}a
+(\frac{1}{8{\cal I}_{1}}-\frac{1}{8{\cal I}_{2}\mu^{2}}(\mu -1))
(a^{\dag}a)^{2} \label{ada} \eea where \bea
\mu&=&(1+\frac{m_{K}^{2}-m_{\pi}^{2}}{m_{0}^{2}})^{1/2}\nonumber\\
m_{0}&=&\frac{N}{4(\chi {\cal I}_{2})^{1/2}} \nonumber \eea and
$a^{\dag}$ is creation operator for constituent strange quarks and
we have ignored the irrelevant creation operator $b^{\dag}$ for
strange antiquarks\cite{kleb94}. Introducing the angular momentum
of the strange quarks $$
\vec{J}_{s}=\frac{1}{2}a^{\dag}\vec{\tau}a, $$
one can rewrite the Hamiltonian (\ref{ada}) as \beq
\tilde{H}=E+\frac{1}{2}m_{\pi}^{2}+\omega a^{\dag}a
+\frac{1}{2{\cal I}_{1}}
(\vec{I}^{2}+2c\vec{I}\cdot\vec{J}_{s}+\bar{c}\vec{J}_{s}^{2}
+\frac{1}{4}) \label{hjs} \eeq where
\begin{eqnarray}
\omega&=&\frac{N}{8{\cal I}_{2}}(\mu -1)  \nonumber \\
c&=&1-\frac{{\cal I}_{1}}{2{\cal I}_{2}\mu}(\mu -1)  \nonumber \\
\bar{c}&=&1-\frac{{\cal I}_{1}}{{\cal I}_{2}\mu^{2}}(\mu -1).
\nonumber
\end{eqnarray}
The Hamiltonian (\ref{hjs}) then yields the structure of the
hyperfine splittings as follows
\begin{eqnarray}
\delta M&=&\frac{1}{2{\cal
I}_{1}}[cJ(J+1)+(1-c)(I(I+1)-\frac{Y^{2}-1}{4}) \nonumber \\ &
&+(1+\bar{c}-2c)\frac{Y^{2}-1}{4}+\frac{1}{4}(1+\bar{c}-c)]
\nonumber
\end{eqnarray}
where $\vec{J}=\vec{I}+\vec{J}_{s}$ is the total angular momentum
of the quarks.

Next, using the Weyl ordering corrected (WOC) energy spectrum
(\ref{hjs}), we easily obtain the hyperfine structure of the
nucleon and delta hyperon masses to yield the soliton energy and
the moment of inertia
\begin{eqnarray}
E&=&\frac{1}{3}(4M_{N}-M_{\Delta})  \nonumber \\ {\cal
I}&=&\frac{3}{2}(M_{\Delta}-M_{N})^{-1}.  \label{masses}
\end{eqnarray}
Substituting the experimental values $M_{N}=939$ MeV and
$N_{\Delta}=1232$
MeV into Eq. (\ref{masses}) and using the expressions for $E$ and ${\cal I}%
_{1}$ given in Eq. (\ref{eni}), one can predict the pion decay constant $%
f_{\pi}$ and the Skyrmion parameter $e$ in the Weyl ordering
corrected rigid rotator approach as follows
\[
f_{\pi}=52.9~{\rm MeV},~~~e=4.88.
\]
With these fixed values of $f_{\pi}$ and $e$, one can then proceed
to evaluate the inertia parameters as follows
\[
{\cal I}_{1}^{-1}= 198~{\rm MeV},~ {\cal I}_{2}^{-1}= 613~{\rm
MeV},~ \chi^{-1}= 257~{\rm MeV},~E=840~{\rm MeV}
\]
to yield the predictions for the values of $c$ and $\bar{c}$
\begin{equation}
c=0.27,~~~\bar{c}=0.23
\end{equation}
which are contained in Table 1, together with the experimental
data and the
standard SU(3) rigid rotator and bound state approach predictions.\footnote{%
Here we have the modified predictions $c=0.22$ and $\bar{c}=0.34$
of the standard rigid rotator without pion mass since the
numerical evaluation for the inertia parameters should be fixed
with the values ${\cal I}_{1}^{-1}= 196~{\rm MeV}$, ${\cal
I}_{2}^{-1}=528~{\rm MeV}$, ${\cal \chi}^{-1}= 182~{\rm MeV}$ and
$E=866~{\rm MeV}$, instead of ${\cal I}_{1}^{-1} =211~ {\rm MeV}$,
${\cal I}_{2}^{-1}=552~{\rm MeV}$, ${\cal \chi}^{-1}=202~{\rm
MeV}$ and $E=862~{\rm MeV}$ which yields $c=0.28$ and
$\bar{c}=0.35$\cite{kleb94}, to be consistent with the parameter
fit $f_{\pi}=64.5~{\rm MeV},~e=5.45$ used in the massless standard
SU(2) Skyrmion\cite{ad}.  Also one notes that the bound state
approach does not include the quartic terms in the kaon field.}
Here one notes that the massless SU(3) Skyrmions have the same
values of $c$ and $\bar{c}$ both in the standard and WOC cases
since the chiral angles are the same in these cases. However, in
the massive Skyrmions the equation of motion for the chiral angle
has an additional term proportional to
$(m_{\pi}/ef_{\pi})^{2}$\cite{ad84} to yield the discrepancies
between the two chiral angles of the standard and WOC cases since
the standard Skyrmion has the values $f_{\pi}=54.0~{\rm MeV}$ and
$e=4.84$ different from the above ones in the massive WOC
Skyrmion.  With these chiral angles and values of $f_{\pi}$ and
$e$, one can obtain different sets of $c$ and $\bar{c}$ in the
massive standard and WOC Skyrmions, which are about $5\%$ improved
with respect to those of the massless Skyrmions as shown in Table
1.

Now we investigate the relations between the Hamiltonian
(\ref{hjs}) and the Berry phases\cite{berry}. In the Berry phase
approach to the SU(3) Skyrmion, the Hamiltonian takes the simple
form\cite{cnd}
\begin{equation}
H^{*}=\epsilon_{K}+\frac{1}{8{\cal
I}_{1}}(\vec{R}^{2}-2g_{K}\vec{R}\cdot
\vec{T}_{K}+g_{K}^{2}\vec{T}_{K}^{2})  \label{h*}
\end{equation}
where $\epsilon_{K}$ is the eigenenergy in the $K$ state, $g_{K}$
is the Berry charge, $\vec{R}$ ($\vec{L}$) is the right (left)
generators of the group $SO(4)\approx SU(2)\times SU(2)$ and
$\vec{T}_{K}$ is the angular
momentum of the "slow" rotation. We recall that $\vec{I}=\frac{\vec{L}}{2} =-%
\frac{\vec{R}}{2}$ and $\vec{L}^{2}=\vec{R}^{2}$ on $S^{3}$.
Applying the BFT scheme to the Hamiltonian (\ref{h*}) we can
obtain the Hamiltonian of the form
\begin{equation}
\tilde{H}^{*}=\epsilon_{K}+\frac{1}{2{\cal
I}_{1}}(\vec{I}^{2}+g_{K}\vec{I}
\cdot\vec{T}_{K}+(\frac{g_{K}}{2})^{2}\vec{T}_{K}^{2}+\frac{1}{4}).
\label{ht*}
\end{equation}
In the case with the relation $\bar{c}=c^{2}$, the Hamiltonian
(\ref{hjs}) is equivalent to $\tilde{H}^{*}$ in the Berry phase
approach where the corresponding physical quantities can be read
off as follows
\begin{eqnarray}
\epsilon_{K}&=&E+\frac{1}{2}\chi m_{\pi}^{2}+\omega a^{\dag}a
\nonumber \\ \vec{T}_{K}&=&\vec{J}_{s}  \nonumber \\ g_{K}&=&2c.
\label{rels}
\end{eqnarray}
The same case with the Hamiltonian (\ref{ht*}) follows from the
quark model and the bound state approach with the quartic terms in
the kaon field neglected. In fact, the strange-strange
interactions in the Hamiltonian (\ref {hjs}) break these relations
to yield the numerical values of $\bar{c}$ in Table 1.

Next, in order to take into account the missing order $N^{0}$
effects, we consider the Casimir energy contributions to the
Hamiltonian (\ref{hjs}). The Casimir energy originated from the
meson fluctuation can be given by the phase shift
formula\cite{mou, park98}
\begin{eqnarray}
E_{C}(\mu)&=&\frac{1}{2\pi}\sum_{i=\pi, K}\{\int_{0}^{\infty}{\rm d}p [-%
\frac{p}{\sqrt{p^{2}+m_{i}^{2}}}(\delta (p)-\bar{a}_{0} p^{3}
-\bar{a}_{1}p) +\frac{\bar{a}_{2}}{\sqrt{p^{2}+\mu^{2}}}]
\nonumber \\
& &-\frac{3}{8}\bar{a}_{0}m_{i}^{4} (\frac{3}{4}+\frac{1}{2}\ln \frac{\mu^{2}%
}{m_{i}^{2}}) +\frac{1}{4}\bar{a}_{1} m_{i}^{2}(1+\ln\frac{\mu^{2}}{m_{i}^{2}%
}) -m_{i}\delta (0)\}  \nonumber \\ & &+\cdots
\end{eqnarray}
where the ellipsis denotes the contributions from the counter
terms and the bound states (if any). Here $\mu$ is the energy
scale and $\delta (p)$ is the phase shift with the momentum $p$
and the coefficients $\bar{a}_{i}$ $(i=0,1,2)$ are defined by the
asymptotic expansion of $\delta ^{\prime} (p)$, namely,
$\delta^{\prime}(p)=3\bar{a}_{0}p^{2}+\bar{a}_{1}-\frac{\bar{a}_{2}}
{p^{2}}+\cdots$.  Even though the Casimir energy correction does
not contribute to the ratios $c$ and $\bar{c}$ since these ratios
are associated with the order $1/N$ piece of the Hamiltonian
(\ref{hjs}), these effects seem to be significant in other
physical quantities such as the H dibaryon mass\cite{kleb96} which
will be studied elsewhere.

\vskip 0.7cm \noindent {\Large {\bf Acknowledgements}}

\vskip 0.7cm One of us (S.T.H.) would like to thank G.E. Brown and
I. Zahed at Stony Brook for discussions and encouragement.  We
would like to thank M. Rho for helpful discussions and for getting
us interested in the Berry phases and Casimir effects.  The
present work is supported by the Ministry of Education, BK21
Project No. D-0055, 1999.

\newpage
\begin{table}[t]
\caption{The values of $c$ and $\bar{c}$ in the bound state and
the standard and Weyl ordering corrected (WOC) rigid rotator
approaches to the massless and massive SU(3) Skyrmions compared
with experimental data.}
\begin{center}
\begin{tabular}{lcc}
\hline Source & $c$ & $\bar{c}$ \\ \hline Bound state, partial &
0.60 & 0.36 \\ Rigid rotator, massless standard & 0.22 & 0.34 \\
Rigid rotator, massless WOC & 0.22 & 0.34 \\ Rigid rotator,
massive standard & 0.26 & 0.23 \\ Rigid rotator, massive WOC &
0.27 & 0.23 \\ Experiment & 0.67 & 0.27 \\ \hline
\end{tabular}
\end{center}
\end{table}
\end{document}